\documentclass[aps,pra,preprint,footinbib,superscriptaddress]{revtex4-1}
\usepackage{graphicx}
\usepackage{indentfirst}
\usepackage{braket}
\usepackage{float}
\usepackage{amsmath}

\usepackage{epsfig,amssymb, amsmath, mathtools}
\newcommand{\bp}{\boldsymbol \rho}
\newcommand{\bpt}{\boldsymbol{\tilde{\rho}}}
\newcommand{\bpp}{\boldsymbol{\rho}_+}
\newcommand{\bpm}{\boldsymbol{\rho}_-}

\newcommand{\bzero}{{\mathbf 0}}
\newcommand{\scP}{{\mathcal P}}
\newcommand{\scPt}{\tilde{\mathcal P}}
\newcommand{\scE}{{\mathcal E}}

\newcommand{\scS}{{\mathcal S}}

\newcommand{\W}[1]{\Omega_{#1}}
\newcommand{\wmod}{\Omega}
\newcommand{\wo}{\omega_0}

\newcommand{\wm}{\omega_-}
\renewcommand{\wp}{\omega_+}
\newcommand{\wpt}{\tilde{\omega}_+}

\newcommand{\nm}{\,\text{nm}}

\newcommand{\cm}{\,\text{cm}}
\newcommand{\m}{\,\text{m}}

\newcommand{\intII}[1]{\int\!{\rm d}^2#1\,}

\newcommand{\intw}[1]{\int\!\frac{{\rm d}#1}{2\pi}\,}

\newcommand{\avg}[1]{\langle #1 \rangle}

\newcommand{\abs}[1]{\left| #1 \right|}
\newcommand{\bracket}[1]{\left[ #1 \right]}
\newcommand{\parens}[1]{\left( #1 \right)}

\newcommand{\BI}{\text{BI}}

\begin{document}
\title{Speckled speckled speckle}

\author{Justin Dove}
\email{dove@mit.edu}
\author{Jeffrey H. Shapiro}
\email{jhs@mit.edu}
\address{Research Laboratory of Electronics, Massachusetts Institute of Technology, Cambridge, MA 02139, USA}
\begin{abstract}
	Speckle is the spatial fluctuation of irradiance seen when coherent light is reflected from a rough surface. It is due to light reflected from the surface's many nooks and crannies accumulating vastly-discrepant time delays, spanning much more than an optical period, en route to an observation point. Although speckle with continuous-wave (cw) illumination is well understood, the emerging interest in non-line-of-sight (NLoS) imaging using coherent light has created the need to understand the higher-order speckle that results from multiple rough-surface reflections, viz., speckled speckle and speckled speckled speckle. Moreover, the recent introduction of phasor-field ($\scP$-field) NLoS imaging---which relies on amplitude-modulated coherent illumination---requires pushing beyond cw scenarios for speckle and higher-order speckle. In this paper, we take first steps in addressing the foregoing needs using a three-diffuser transmissive geometry that is a proxy for three-bounce NLoS imaging.  In the small-diffusers limit, we show that the irradiance variances of cw and modulated $n$th-order speckle coincide and are $(2^n-1)$-times those of ordinary (first-order) speckle.  The more important case for NLoS imaging, however, involves extended diffuse reflectors.  For our transmissive geometry with extended diffusers, we treat third-order cw speckle and first-order modulated speckle.  Our results there imply that speckle is unlikely to impede successful operation of coherent-illumination cw imagers, and they suggest that the same might be true for $\scP$-field imagers.
\end{abstract}

\maketitle

\section{Introduction}

When continuous-wave (cw) laser light that has been diffusely reflected by a rough surface illuminates an observation plane some distance away, a speckle pattern is visible in the measured irradiance. This speckle is due to wave-optical interference between reflections from independent wavelength-scale surface patches that vary in height by many optical wavelengths. This phenomenon, which we call ordinary (first-order) speckle, is well studied~\cite{goodman-speckle}. The irradiance produced by diffuse reflection is exponentially distributed at any observation point, and thus has a variance equal to its squared mean. The irradiance's covariance function for diffuse reflection from an extended surface is also well understood, as are limited properties of the second-order speckle seen when the speckle pattern from a first diffuse reflection is reflected from a second rough surface and observed, as speckled speckle, on a new observation plane. However, with the growing interest in non-line-of-sight (NLoS) imaging, colloquially referred to as ``seeing around corners'', there is now an unfulfilled need to understand higher-order speckle effects, e.g., speckled speckled speckle, such as results from laser light sequentially reflecting off three rough surfaces. Moreover, the advent of phasor-field ($\scP$-field) NLoS imagers~\cite{Reza2018,Dove2019,Teichman2019}---which rely on amplitude-modulated coherent illumination---dictates that the preceding speckle questions be addressed for amplitude-modulated as well as for cw illumination. 

In this paper, we take first steps in addressing the preceding issues using a paraxial, scalar-wave, transmissive geometry---which is a proxy for three-bounce NLoS imaging---that we have employed in our earlier treatments of $\scP$-field imaging with quasimonochromatic coherent illumination~\cite{Dove2019,Dove2020}.   After some preliminaries, which allow us to obtain the complete statistics of cw and modulated speckled speckled speckle in the small-diffusers limit, we begin our analyses in earnest with third-order speckle for cw illumination of extended diffusers. There, although small-diffuser speckled speckled speckle is seven times stronger than ordinary (first-order) speckle---i.e., its irradiance variance is seven times its squared mean---we find that cw speckled speckled speckle is highly mitigated by the geometry of the problem. In particular, our closed-form expression for the irradiance covariance of cw third-order speckle proves that the geometry of typical NLoS imaging scenarios reduces that third-order speckle to the ordinary cw speckle produced by the final diffuser.  Furthermore, the speckle fluctuations that remain will be suppressed in power collection over any reasonable detector area. To quantify the impact of those residual power fluctuations, we evaluate the signal-to-noise ratio (SNR) for direct detection and show that power fluctuations act as an excess noise---above the fundamental shot-noise limit---that sets a maximum attainable SNR.  For typical parameter values, we find this saturation SNR to be quite generous and thus conclude that the impact of third-order speckle is unlikely to be significant in NLoS imaging with cw coherent illumination.  

Next, we move on to the speckle produced by the modulated illumination of extended diffusers, as used in $\scP$-field imaging. Here, our analysis is limited to first-order speckle. We establish an upper bound on the zero-frequency component of the speckle when the initial illumination is space-time factorable finding that, at worst, such speckle is of ordinary strength. This result is seemingly at odds with Teichman's analysis~\cite{Teichman2019} for factorable, single-frequency modulation, which finds the modulation-frequency-component speckle to be \emph{stronger} than ordinary speckle. To resolve the apparent discrepancy, we analyze a single-frequency-modulation limiting case in our $\scP$-field framework and show that it recovers both Teichman's result for the modulation-frequency speckle \emph{and} our upper bound for the zero-frequency speckle. Then, using realistic parameter values for NLoS imaging scenarios, we conclude that the speckle enhancement effect reported by Teichman is likely to be minimal. Moving further, we analyze the first-order-speckle size for the modulated case and find that it is close to that of the cw case's first-order speckle.  If these similarities---for both speckle strength and speckle size---between modulated and cw speckle from extended diffusers persist in their second-order and third-order cases, then the adverse effects of speckled speckled speckle on $\scP$-field NLoS imaging may be inconsequential.  

\section{Preliminaries}
As developed in our earlier analysis of $\scP$-field imaging~\cite{Dove2019,thesis,Dove2020}, we use paraxial, scalar-wave optics in a transmissive geometry that serves as a proxy for a typical reflective, three-bounce NLoS geometry. The light at each plane is characterized by its baseband complex-field envelope $E_z(\bp_z,t)$, which modulates an optical carrier of frequency $\wo$ to produce a ${\rm W}^{1/2}/{\rm m}$-units optical field $\text{Re}[E_z(\bp_z,t)e^{-i\wo t}]$, where $\bp_z$ is the 2D transverse spatial coordinate in the plane indicated by $z$. For cw speckle $E_z(\bp_z,t) = E_z(\bp_z)$ will have no time dependence, whereas for modulated speckle $E_z(\bp_z,t)$ will have bandwidth $\Delta\omega \ll \omega_0$.  In both cases, $I_z(\bp_z,t) \equiv |E_z(\bp_z,t)|^2$ will be the short-time-average (STA) irradiance at the $z$-plane~\cite{footnote1}.

\begin{figure}[hbt]
	\centering
	\includegraphics[width=4.5in]{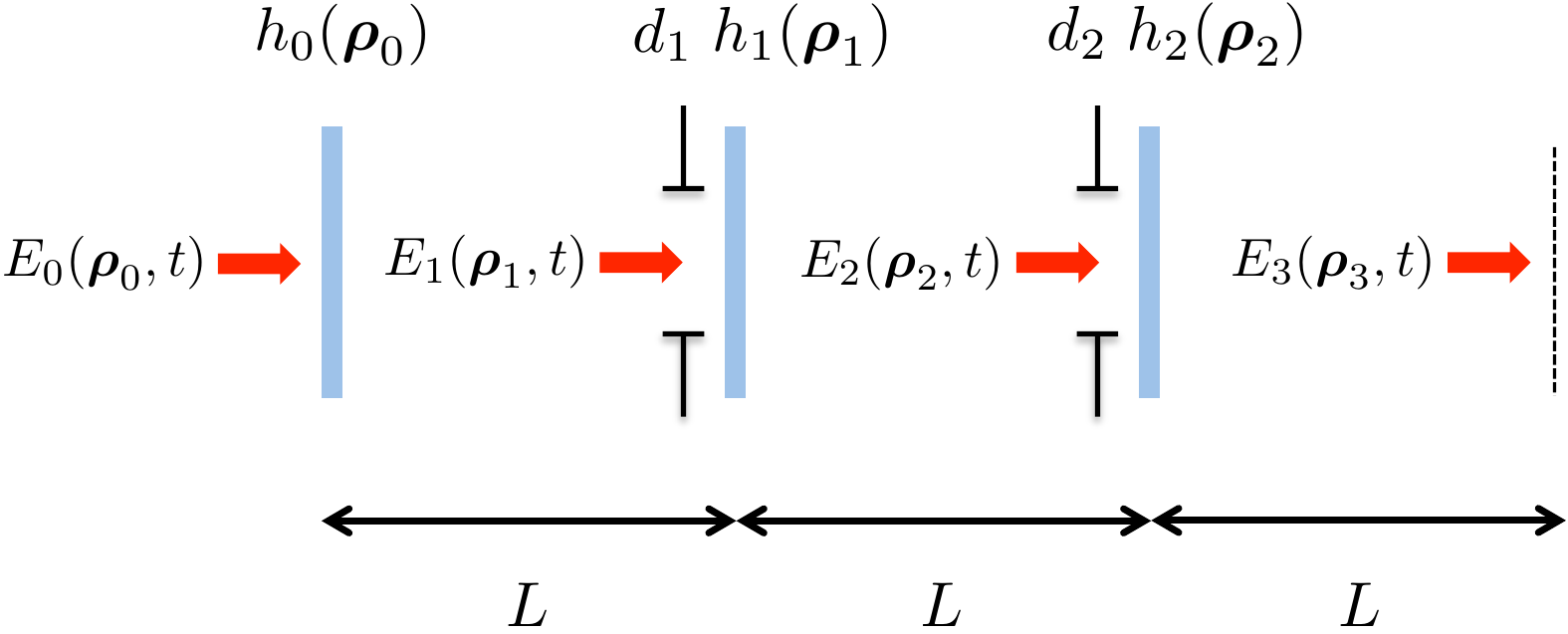}
	\caption{Geometry for third-order speckle analysis. Thin blue rectangles represent idealized, thin diffusers. The black frames in front of the diffusers in planes 1 and 2 represent Gaussian pupils that capture the essence of the target and visible-wall sizes, respectively. The dashed line represents the detection plane.
		\label{fig:geo}}
\end{figure}

The geometry for our third-order speckle analysis is depicted in Fig.~\ref{fig:geo}. Coherent space-time factorable illumination, $E_0(\bp_0,t) = E_0(\bp_0)S(t)$ with 
\begin{align}
	E_0(\bp_0) = \sqrt{I_0}\,e^{-4\abs{\bp_0}^2/d_0^2},
	\label{eq:speckle-input}
\end{align}
and $S(t) = 1$ for cw illumination, is incident at plane~0, which contains a diffuser with thickness profile $h_0(\bp_0)$.  In the standard NLoS imaging configuration, this diffuser represents the visible wall at which reflection into the hidden space occurs. Plane~1 contains a second diffuser, whose thickness profile is $h_1(\bp_1)$ and whose size is modeled by a Gaussian pupil with $e^{-1}$-field-attenuation-diameter $d_1$.  This diffuser represents a finite-sized, planar, diffuse target in the hidden scene, where, for simplicity, we have ignored any albedo variations across the target. Plane~2 contains a final diffuser, whose thickness profile is $h_2(\bp_2)$ and whose finite size is modeled by a Gaussian pupil with $e^{-1}$-field-attenuation-diameter $d_2$.  In the NLoS scenario, it represents the visible wall where light reflects back to the imager.  That imager's entrance pupil lies in plane~3.  Note that plane~2's finite pupil enables us to obtain convergent paraxial-regime results for the variance of third-order cw speckle.  A finite pupil at plane~0 is not needed for that purpose, because the initial illumination is self-limited to within that wall's boundaries.  The distances between planes~0, 1, 2, and 3 are all $L$, a choice made for convenience rather than necessity.

\subsection{Basic principles \label{basics}}
All the analysis to follow rests on four basic principles:  Fresnel diffraction for monochromatic light; the van Cittert--Zernike theorem for propagating the mutual coherence function (MCF) of spatially-incoherent light; the central limit theorem for sums of large numbers of independent random variables; and the law of iterated expectation.  

To see how these principles come into play in the cw case, we start with how the initial illumination, $E_0(\bp_0)$ from Eq.~(\ref{eq:speckle-input}), first passes through the plane-0 diffuser to become $E_0'(\bp_0)$ and then diffracts over a length $L$ free-space path to become the illumination, $E_1(\bp_1)$, at plane~1.  We have that
\begin{equation}
E'_0(\bp_0) = E_0(\bp_0)e^{i\omega_0h_0(\bp_0)/c},
\label{diffuser_effect}
\end{equation}
where $c$ is light speed and we have ignored the diffuser's refractive index.  The essence of $\scP$-field imaging~\cite{Reza2018} is that the diffusers are rough at the optical wavelength, $\lambda_0 = 2\pi c/\omega_0$, but smooth at the modulation wavelength, $\Delta\lambda = 2\pi c/\Delta\omega$ \cite{footnote2}.  Our previous work~\cite{Dove2019,Dove2020,thesis} enforced this behavior by taking the $\{h_n(\bp_n): n = 0,1,2\}$ to be statistically-independent~\cite{footnote3}, identically-distributed, zero-mean Gaussian random processes whose 
covariance function
\begin{equation}
K_{hh}(\bp_n,\bpt_n) \equiv \langle h_n(\bp_n)h_n(\bpt_n)\rangle = \sigma_h^2 e^{-|\bp_n-\bpt_n|^2/\rho_h^2},
\end{equation}
has a standard deviation $\sigma_h$ satisfying $\lambda_0 \ll \sigma_h \ll \Delta\lambda$ and a correlation length $\rho_h$ satisfying $\rho_h \sim \lambda_0$.  For the cw case, with $\langle \cdot\rangle_0$ denoting ensemble averaging over $h_0(\bp_0)$, these statistics imply that 
\begin{equation}
\langle E'_0(\bp_0)\rangle_0 = E_0(\bp_0) e^{-\wo^2\sigma_h^2/2c^2} \approx 0, 
\end{equation}
\begin{equation}
\langle E'_0(\bp_0)E'_0(\tilde{\bp}_0)\rangle_0  = E_0(\bp_0)E_0(\bpt_0) \exp\!\left[-\wo^2\sigma_h^2\parens{1+e^{-|\bp_0-\bpt_0|^2/\rho_h^2}}/c^2\right] \approx 0,
\end{equation}
and
\begin{align}
\langle E'_0(\bp_0) E^{\prime *}_0(\tilde{\bp}_0)\rangle_0 &=  E_0(\bp_0)E^*_0(\tilde{\bp}_0)\exp\!\left[-\wo^2\sigma_h^2\parens{1-e^{-|\bp_0-\bpt_0|^2/\rho_h^2}}/c^2\right] \\
&\approx E_0(\bp_0)E^*_0(\tilde{\bp}_0)\lambda_0^2\delta(\bp_0-\tilde{\bp}_0),
\label{impulseApprox}
\end{align}
where $\delta(\cdot)$ is the unit impulse and the approximation uses the fact that the smallest correlation area for a wavelength-$\lambda_0$ propagating wave is $\sim$$\lambda_0^2$~\cite{footnoteA}.

Fresnel diffraction at frequency $\omega_0$ now gives
\begin{equation}
E_1(\bp_1) = \frac{e^{i\omega_0L/c}}{i\lambda_0L}\int\!{\rm d}^2\bp_0\,E'_0(\bp_0)e^{i\omega_0 |\bp_1-\bp_0|^2/2cL},
\label{cwFresnel}
\end{equation}
and we can use this result, in conjunction with the approximation in (\ref{impulseApprox}), to obtain
\begin{equation}
\langle E_1(\bp_1)E_1^*(\tilde{\bp}_1)\rangle_0 = \frac{e^{i\omega_0(|\bp_1|^2-|\tilde{\bp}_1|^2)/2cL}}{L^2}\int\!{\rm d}^2\bp_0\,|E_0(\bp_0)|^2e^{i\omega_0(\tilde{\bp}_1-\bp_1)\cdot \bp_0/cL},
\label{cwMCF}
\end{equation}
via the van Cittert--Zernike theorem \cite{goodman-speckle}.  This result, together with Eq.~(\ref{eq:speckle-input}), immediately proves that the cw case's diffuser-averaged STA irradiance at plane~1 is independent of $\bp_1$ and given by
$\langle I_1\rangle \equiv \langle I_1(\bp_1)\rangle_0 = \pi d_0^2I_0/8L^2$~\cite{footnote4}.  Now, by using Eq.~(\ref{diffuser_effect}) in Eq.~(\ref{cwFresnel}), the central limit theorem tells us that $E_1(\bp_1)$ will be a zero-mean, complex-valued, Gaussian random process that is completely characterized by the MCF from Eq.~(\ref{cwMCF}).  This result then implies that $I_1(\bp_1)$ is exponentially distributed and so has ${\rm Var}[I_1(\bp_1)] = \langle I_1\rangle^2$.

The final principle our analysis will need---iterated expectation---comes in at this point.  \emph{Conditioned} on knowledge of $|E_1(\bp_1)|^2$, the procedure we have just employed can be used to show that $E_2(\bp_2)$ is a zero-mean, complex-valued, Gaussian random process that is completely characterized by
\begin{equation}
\langle E_2(\bp_2)E_2^*(\tilde{\bp}_2)\rangle_1 = \frac{e^{i\omega_0(|\bp_2|^2-|\tilde{\bp}_2|^2)/2cL}}{L^2}\int\!{\rm d}^2\bp_1\,|E_1(\bp_1)|^2e^{-8|\bp_1|^2/d_1^2}e^{i\omega_0(\tilde{\bp}_2-\bp_2)\cdot \bp_1/cL},
\label{MCF2Condx}
\end{equation}
where we have used $E_1'(\bp_1) = E_1(\bp_1)e^{-4|\bp_1|^2/d_1^2}e^{i\omega_0h_1(\bp_1)/c}$ for the field emerging from the diffuser at plane~1 and $\langle \cdot \rangle_1$ denotes ensemble averaging over that diffuser.  Iterated expectation now gives us 
\begin{equation}
\langle E_2(\bp_2)E_2^*(\tilde{\bp}_2)\rangle_{0,1} =  \frac{e^{i\omega_0(|\bp_2|^2-|\tilde{\bp}_2|^2)/2cL}}{L^2}\int\!{\rm d}^2\bp_1\,\langle |E_1(\bp_1)|^2\rangle_0\, e^{-8|\bp_1|^2/d_1^2}e^{i\omega_0(\tilde{\bp}_2-\bp_2)\cdot \bp_1/cL},
\label{MCF2}
\end{equation}
for the \emph{unconditional} (ensemble averaged over the plane-0 and plane-1 diffusers) MCF of $E_2(\bp_2)$. Hence we get $\langle I_2\rangle \equiv \langle I_2(\bp_2) \rangle_{0,1} = \pi d_1^2 \langle I_1\rangle/8L^2$ for the unconditional, diffuser-averaged STA irradiance at plane~2.   By now it should be clear that we can pursue a similar argument to that just completed and show that $\langle I_3\rangle \equiv \langle I_3(\bp_3)\rangle_{0,1,2} = \pi d_2^2\langle I_2\rangle/8L^2$.

We can obtain a further result for cw speckled speckle in the small-diffuser limit, wherein ($d_0d_1/4\lambda_0 L)^2 \ll 1$.   Small diffusers are especially interesting because they give rise to large speckles that prevent speckle-fluctuation suppression via spatial averaging, i.e., they imply worst-case signal-to-noise ratio behavior, as we will see later.  In particular, within this regime we have that
\begin{equation}
|E_1(\bp_1)|^2 e^{-8|\bp_1|^2/d_1^2} \approx |E_1({\bf 0})|^2 e^{-8|\bp_1|^2/d_1^2}.
\label{smallDiffuse1}
\end{equation} 
Consequently, the \emph{unconditional} probability density function (pdf) for $I_2(\bp_2) = |E_2(\bp_2)|^2$ in the small-diffuser limit, i.e., the pdf for cw speckled speckle in that regime, is~\cite{goodman-speckle}
\begin{align}
p_{I_2}(\mathcal{I}_2) &= \int_0^\infty\!{\rm d}\mathcal{I}_1\, \frac{\exp(-\mathcal{I}_1/\langle I_1\rangle)}{\langle I_1\rangle} \frac{\exp[-\mathcal{I}_2/(\pi d_1^2\mathcal{I}_1/8L^2)]}{(\pi d_1^2\mathcal{I}_1/8L^2)}u(\mathcal{I}_2) 
= \left[2K_0\!\parens{2\sqrt{\mathcal{I}_2/\langle I_2\rangle}}/\langle I_2\rangle\right]\!u(\mathcal{I}_2),
\end{align}
where $K_0(\cdot)$ is the zeroth-order modified Bessel function of the second kind and $u(\cdot)$ is the unit-step function.

The results in this subsection can be found in Goodman's monograph~\cite{goodman-speckle}.  We have reviewed them for two reasons.  First, in Sec.~\ref{smalldiffusers}, they will let us analyze both cw \emph{and} modulated speckled speckle speckle in the small-diffusers limit.  Second, in Secs.~\ref{cwExtended} and \ref{ModExtended}, respectively, they will be generalized to treat cw third-order speckle from extended diffusers and modulated first-order speckle from an extended diffuser.  

\subsection{Third-order speckle in the small-diffusers regime \label{smalldiffusers}}
\begin{figure}[hbt]
	\hspace*{0.9in}
	\includegraphics[width=2.75in]{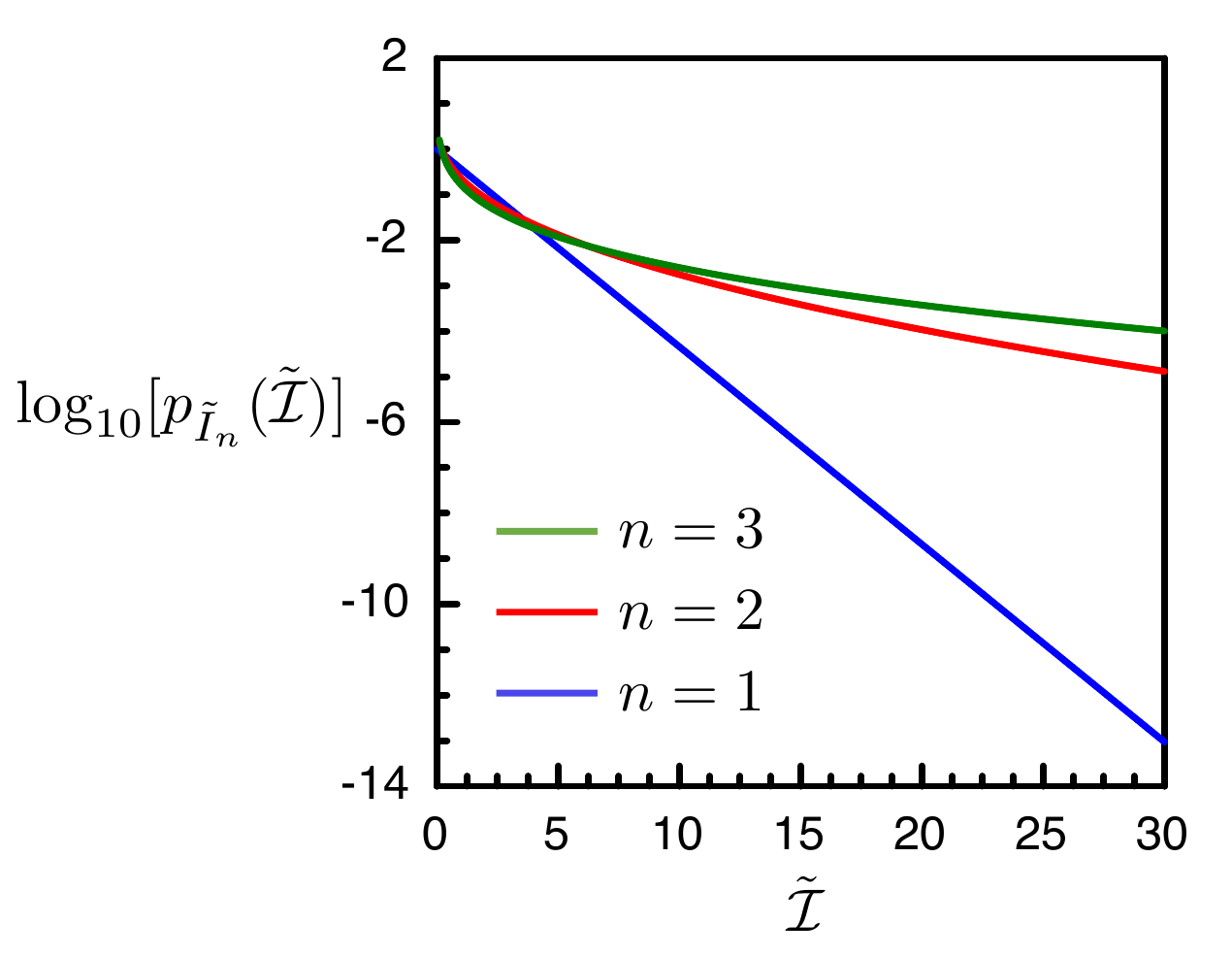}
	\caption{Logarithmic plots of the pdfs for $\tilde{I}_n \equiv I_n({\bf 0})/\langle I_n\rangle$, the normalized $n$th-order speckle in the small-diffusers limit.
		\label{specklePDFs}}
\end{figure}

Guided by the small-diffuser result for cw second-order speckle, and motivated by the desire to quantify worst-case speckle behavior for three-bounce NLoS imaging, let us consider cw third-order speckle when $(d_0d_1/4\lambda_0L)^2 \ll 1$ \emph{and} $(d_1d_2/4\lambda_0L)^2 \ll 1$, i.e., when the Fresnel-number products for propagation between planes~0 and 1 and between planes~1 and 2 are both very small.  The work from the previous subsection immediately shows us that \begin{equation}
|E_2(\bp_2)|^2 e^{-8|\bp_2|^2/d_2^2} \approx |E_2({\bf 0})|^2 e^{-8|\bp_2|^2/d_2^2},
\label{smallDiffuse2}
\end{equation}
so that, conditioned on knowledge of $|E_2({\bf 0})|^2$, we have that $E_3(\bp_3)$ is a zero-mean, complex-valued, Gaussian random process that is completely characterized by its conditional MCF,
\begin{equation}
\langle E_3(\bp_3)E_3^*(\tilde{\bp}_3)\rangle_2 = \frac{e^{i\omega_0(|\bp_3|^2-|\tilde{\bp}_3|^2)/2cL}}{L^2}|E_2({\bf 0})|^2\int\!{\rm d}^2\bp_2\,e^{-8|\bp_2|^2/d_2^2}e^{i\omega_0(\tilde{\bp}_3-\bp_3)\cdot \bp_2/cL}.
\label{MCF3Condx}
\end{equation}  The unconditional pdf for $I_3(\bp_3) = |E_3(\bp_3)|^2$ in the small-diffusers limit, i.e., the pdf for cw speckled speckled speckle in that regime, is therefore
\begin{equation}
p_{I_3}(\mathcal{I}_3) = \int_0^\infty\!{\rm d}\mathcal{I}_2\, \frac{\displaystyle 2K_0\!\parens{2\sqrt{\mathcal{I}_2/\langle I_2\rangle}}}{\displaystyle \langle I_2\rangle}\frac{\exp[-\mathcal{I}_3/(\pi d_2^2\mathcal{I}_2/8L^2)]}{(\pi d_2^2\mathcal{I}_2/8L^2)}u(\mathcal{I}_3).
\end{equation}
Recourse to integral tables yields
\begin{equation}
p_{I_3}(\mathcal{I}_3) = [G_{0,3}^{3,0}(\mathcal{I}_3/\langle I_3\rangle \vert 0,0,0)/\langle I_3\rangle]u(\mathcal{I}_3),
\end{equation}
where $G_{0,3}^{3,0}(\cdot\vert 0,0,0)$ is a Meijer $G$-function~\cite{Meijer}.  In Fig.~\ref{specklePDFs} we have plotted the pdfs of the normalized irradiances $\tilde{I}_n \equiv I_n({\bf 0})/\langle I_n\rangle$ for $n = 1, 2, 3$, which show the increasing randomness that occurs in progressing from speckle to speckled speckle to speckled speckled speckle.  Indeed, more pdf iterations and properties of the Meijer $G$-function can be used to show that the unconditional pdf for $n$th-order speckle in the small-diffusers regime is~\cite{thesis}
\begin{equation}
p_{I_n}(\mathcal{I}_n) = [G_{0,n}^{n,0}(\mathcal{I}_n/\langle I_n\rangle \vert 0,\ldots,0)/\langle I_n\rangle]u(\mathcal{I}_n),
\end{equation}
where $\langle I_n\rangle = \Pi_{m=0}^{n-1}(\pi d_m^2/8L^2)I_0$.  The normalized variance then turns out to be ~\cite{thesis}
\begin{equation}
{\rm NVar}_{I_n} \equiv {\rm Var}[I_n({\bf 0})]/\langle I_n\rangle^2 = (2^n-1).
\end{equation}

Interestingly, this subsection's results for high-order cw speckle in the small-diffusers limit have immediate translations into corresponding results for high-order modulated speckle in that regime.  In particular, with our assumption of quasimonochromatic, space-time factorable, modulated illumination, the small-diffuser assumption $(d_0d_1/4\lambda_0L)^2 \ll 1$ leads to 
\begin{align}
E_1(\bp_1,t)e^{-4|\bp_1|^2/d_1^2} &= e^{i\omega_0L/c}\left(\int\!{\rm d}^2\bp_0\,\frac{E'_0(\bp_0)}{i\lambda_0L}\right)S(t-L/c)e^{-4|\bp_1|^2/d_1^2}\\[.05in]
& = E_1({\bf 0})S(t-L/c)e^{-4|\bp_1|^2/d_1^2},
\end{align}
where $E_1({\bf 0})$ is the cw-illumination complex envelope whose squared magnitude appears in Eq.~(\ref{smallDiffuse1}).  Similarly, because we have just shown that $E_1(\bp_1,t)$ for quasimonochromatic, space-time factorable, initial illumination is itself quasimonochromatic and space-time factorable when $(d_0d_1/4\lambda_0L)^2 \ll 1$, we have that the additional small-diffuser assumption $(d_1d_2/4\lambda_0L)^2 \ll 1$ leads to 
\begin{align}
E_2(\bp_2,t)e^{-4|\bp_2|^2/d_2^2} &= e^{i\omega_0L/c}\left(\int\!{\rm d}^2\bp_1\,\frac{E'_1(\bp_1)}{i\lambda_0L}\right)S(t-2L/c)e^{-4|\bp_2|^2/d_2^2}\\[.05in]
& = E_2({\bf 0})S(t-2L/c)e^{-4|\bp_2|^2/d_2^2},
\end{align}
where $E_2({\bf 0})$ is the cw-illumination complex envelope whose squared magnitude appears in Eq.~(\ref{smallDiffuse2}).  So, except for their having time-delayed temporal modulations $|S(t-2L/c)|^2$ and $|S(t-3L/c)|^2$, the behaviors of second-order and third-order modulated speckle in the small-diffusers regime are identical to what we found for their cw counterparts.  

\section{Third-order cw speckle from extended diffusers \label{cwExtended}}
First-order speckle has long been an issue for line-of-sight laser radars.  For a rough-surfaced target, the single-pulse, single-pixel SNR of a heterodyne-detection laser radar asymptotes to a saturation SNR of 1---set by first-order speckle---with increasing target-return strength~\cite{Shapiro1981}.   Direct-detection laser radars are largely immune to first-order speckle because each pixel is configured to contain sufficient speckles to average out their individual fluctuations without unduly compromising spatial resolution. NLoS laser imagers could potentially suffer third-order speckle's seven-fold increased fluctuation strength that prevails with small diffuse reflectors.  If unabated, this increase would result in a saturation SNR of 1/7.  Whether or not such will be the case requires understanding the statistics of third-order speckle from extended targets, which is the case of interest for NLoS laser imagers.  That task, for the cw case, is this section's mission.  We begin by relating direct detection's saturation SNR to third-order speckle's irradiance statistics. 

Suppose plane~0 in Fig.~\ref{fig:geo} is illuminated with cw light from Eq.~(\ref{eq:speckle-input}) and that a direct-detection system integrates the optical power transmitted through a diameter-$D$ circular pupil in plane~3 over the time interval $0\le t\le T$.  We will neglect technical noises, e.g., thermal noise, and normalize the detector's output to represent the number of detected photons, $N$, in that time interval.  By the conditional Poissonian nature of photon-counting statistics for randomized laser light~\cite{semiclassical}, we have that the resulting SNR is
\begin{equation}
{\rm SNR} \equiv \frac{\langle N\rangle^2}{{\rm Var}(N)} = \frac{(\eta \langle P_3\rangle T/\hbar\omega_0)^2}{\eta \langle P_3\rangle T/\hbar\omega_0 + \eta^2{\rm Var}(P_3)T^2/(\hbar\omega_0)^2},
\label{SNR1}
\end{equation}
where $\eta$ is the detector's quantum efficiency, $\hbar\omega_0$ is the photon energy, and 
\begin{equation}
P_3 = \int_{|\bp_3|\le D/2}\!{\rm d}^2\bp_3\,I_3(\bp_3)
\end{equation}
is the detected power.  The first term in the SNR's denominator is due to shot noise---the fundamental noise of semiclassical photodetection~\cite{semiclassical}, which is \emph{always} present---and the second term in that denominator is the excess noise associated with randomness in the detector's illumination.
We can rewrite Eq.~(\ref{SNR1}) as
\begin{equation}
{\rm SNR} = \frac{{\rm SNR}_{\rm sat}}{{\rm SNR}_{\rm sat}/\langle N\rangle + 1},
\label{SNR2}
\end{equation}
where the saturation SNR, 
\begin{equation}
{\rm SNR}_{\rm sat} \equiv \frac{\langle P_3\rangle^2}{{\rm Var}(P_3)} = \frac{(\pi D^2\langle I_3\rangle/4)^2}{\int_{|\bp_3|\le D/2}\!{\rm d}^2\bp_3\int_{|\tilde{\bp}_3|\le D/2}\!{\rm d}^2\tilde{\bp}_3\,{\rm Covar}[I_3(\bp_3),I_3(\tilde{\bp}_3)]},
\end{equation}
is the maximum achievable SNR, and it is only approached (from below) as $\langle N\rangle \rightarrow \infty$.
From Sec.~\ref{basics} we have that $\langle I_3\rangle = \pi^3d_0^2d_1^2d_2^2 I_0/512 L^6$.  All that remains, before we can evaluate ${\rm SNR}_{\rm sat}$, is to find plane 3's normalized irradiance covariance,
\begin{equation}
{\rm NCovar}_{I_3}(\bp_3-\tilde{\bp}_3) \equiv \frac{{\rm Covar}[I_3(\bp_3),I_3(\tilde{\bp}_3)]}{\langle I_3\rangle^2},
\end{equation}
where we are anticipating its being spatially homogeneous, as indeed will turn out to be the case.
We will find this normalized irradiance covariance in the next subsection, using Gaussian moment factoring and iterated expectation.   Along the way we will get the normalized covariances for $I_1(\bp_1)$ and $I_2(\bp_2)$, whose behaviors aid our understanding of how extended diffusers mitigate high-order speckle.

\subsection{Irradiance covariance of cw third-order speckle \label{irradcovar}}
The normalized covariance of plane~1's irradiance is easily obtained.  We know that $E_1(\bp_1)$ is a zero-mean, complex-valued, Gaussian random process that is completely characterized by the MCF from Eq.~(\ref{cwMCF}).  Gaussian moment factoring gives us
\begin{align}
{\rm Covar}[I_1(\bp_1),I_1(\tilde{\bp}_1)]  &= \langle |E_1(\bp_1)|^2|E_1(\tilde{\bp}_1)|^2\rangle_0-\langle I_1\rangle^2 \\[.05in]
&= |\langle E_1(\bp_1)E_1^*(\tilde{\bp}_1)\rangle_0|^2 \\[.05in]
&=\left|\frac{e^{i\omega_0(|\bp_1|^2-|\tilde{\bp}_1|^2)/2cL}}{L^2}\int\!{\rm d}^2\bp_0\,|E_0(\bp_0)|^2e^{i\omega_0(\tilde{\bp}_1-\bp_1)\cdot \bp_0/cL}\right|^2,
\end{align}
which is spatially homogeneous, as presumed earlier.  Plane 1's normalized irradiance covariance is then found to be
\begin{align}
	{\rm NCovar}_{I_1}(\bp_1-\bpt_1) =   e^{-4 \W{01} \abs{\bp_1-\bpt_1}^2 / d_1^2},
	\label{eq:sp1}
\end{align}
where $\Omega_{01} \equiv (\pi d_0d_1/4\lambda_0L)^2$.

Proceeding now toward obtaining plane 2's irradiance covariance, we start from $E_2(\bp_2)$'s being---conditioned on knowledge of $E_1(\bp_1)$---a zero-mean, complex-valued, Gaussian random process that is completely characterized by its conditional MCF from Eq.~(\ref{MCF2Condx}).  Gaussian moment factoring now gives us $I_2(\bp_2)$'s conditional correlation function,
\begin{align}
	\avg{I_2(\bp_2)I_2(\bpt_2)}_{1} =& \avg{E_2(\bp_2)E_2^*(\bp_2)E_2(\bpt_2)E_2^*(\bpt_2)}_1 \\
	=& \avg{I_2(\bp_2)}_1 \avg{I_2(\bpt_2)}_1
		+ |\avg{E_2(\bp_2)E_2^*(\bpt_2)}_1|^2.
\end{align}
Using Fresnel propagation, these terms expand to give
\begin{align}
	\avg{I_2(\bp_2)I_2(\bpt_2)}_{1} =& \frac{1}{L^4} \Bigg[ \intII{\bp_1}\intII{\bpt_1} I_1(\bp_1)I_1(\bpt_1) e^{-8 (\abs{\bp_1}^2 + \abs{\bpt_1}^2) / d_1^2} \nonumber\\&\times \parens{1 + e^{-i\omega_0(\bp_1-\bpt_1)\cdot(\bp_2-\bpt_2) / c L}} \Bigg].
\end{align}
Now, using the law of iterated expectation and taking advantage of the linearity of expectation, averaging over the first-diffuser's statistics yields
\begin{align}
	\avg{I_2(\bp_2)I_2(\bpt_2)}_{0,1} =& \frac{1}{L^4} \Bigg[ \intII{\bp_1}\intII{\bpt_1} \avg{I_1(\bp_1)I_1(\bpt_1)}_0 e^{-8 (\abs{\bp_1}^2 + \abs{\bpt_1}^2) / d_1^2} \nonumber\\&\times \parens{1 + e^{-i\omega_0(\bp_1-\bpt_1)\cdot(\bp_2-\bpt_2) / c L}} \Bigg],
	\label{eq:wss}
\end{align}
for $I_2(\bp_2)$'s unconditional correlation function.   Using this result we get
\begin{align}
	{\rm NCovar}_{I_2}(\bp_2-\bpt_2) =&  e^{-4 \W{01} \abs{\bp_2-\bpt_2}^2 / d_0^2}   + \frac{1}{1+\W{01}} \parens{1 +  e^{-4\W{01} \abs{\bp_2-\bpt_2}^2 / d_0^2\parens{1+\W{01}}}},
	\label{eq:sp2}
\end{align}
which again is spatially homogeneous.  This normalized covariance has interesting behavior with an interesting interpretation.  When $\Omega_{01} \gg 1$,  it reduces to 
\begin{equation}
{\rm NCovar}_{I_2}(\bp_2-\bpt_2) =  e^{-(\pi d_1/2\lambda_0L)^2 \abs{\bp_2-\bpt_2}^2 }
\end{equation}
which resembles the normalized covariance for first-order speckle, cf.
\begin{equation}
{\rm NCovar}_{I_1}(\bp_1-\bpt_1) =  e^{-(\pi d_0/2\lambda_0L)^2 \abs{\bp_1-\bpt_1}^2},
\end{equation}
which follows from Eq.~(\ref{eq:sp1}).  This is not an accidental coincidence.  When $\Omega_{01} \gg 1$, the speckle size in $I_1(\bp_1)$ is much \emph{smaller} than $d_1$.  Moreover, $I_2(\bp_2)$ is conditionally exponential, given $I_1(\bp_1)$, with conditional mean $\langle I_2\rangle_1 = \int\!{\rm d}^2\bp_1\,I_1(\bp_1)e^{-8|\bp_1|^2/d_1^2}/L^2$.  By the law of large numbers, we have that $\int\!{\rm d}^2\bp_1\,I_1(\bp_1)e^{-8|\bp_1|^2/d_1^2}/L^2 \approx \pi d^2_1\langle I_1\rangle/8L^2$, equivalently $\langle I_2(\bp_2)\rangle_1 \approx \langle I_2\rangle$, because of speckle averaging over the plane-1 pupil.  Furthermore, this means we can take $E_2'(\bp_2) \equiv E_2(\bp_2)e^{i\omega_0h_2(\bp_2)/c}$ to be a zero-mean, complex-valued, Gaussian random process insofar as calculating the statistics of $E_3(\bp_3)$ is concerned, i.e., $\Omega_{01} \gg 1$ has totally suppressed the speckle generated in propagation from plane~0 to plane~1 insofar as evaluating the speckle incurred in propagating from plane~1 to plane~2.

Putting aside, for now, the $\Omega_{01} \gg 1$ condition and its consequences, it should be clear that even without that condition, we can proceed with an iterated-expectation procedure to find ${\rm NCovar}_{I_3}(\bp_3-\bpt_3)$.  For the sake of brevity, we will omit the details and just give the final answer:
\begin{align}
	{\rm NCovar}_{I_3}(\bp_3-\bpt_3)  =&  e^{-4 \W{12} \abs{\bp_3-\bpt_3}^2 / d_1^2}  + \frac{1}{1+\W{01}}\parens{1 +  e^{-4 \W{12} \abs{\bp_3-\bpt_3}^2 / d_1^2}} \nonumber\\&+ \frac{1}{1+\W{12}} \parens{ 1 
	  +  e^{-4\W{12} \abs{\bp_3-\bpt_3}^2 / d_1^2\parens{1+\W{12}}}}
	\nonumber\\&+ \frac{1}{1+\W{01}+\W{12}} \parens{1  +  e^{-4\W{12}\parens{1+\W{01}} \abs{\bp_3-\bpt_3}^2 / d_1^2\parens{1+\W{01}+\W{12}}}},
	\label{eq:sp3}
\end{align}
where $\Omega_{12} \equiv (\pi d_1d_2/4\lambda_0L)^2$.  

When $\Omega_{01} \gg 1$ \emph{and} $\Omega_{12} \gg 1$, the preceding normalized covariance becomes
\begin{equation}
{\rm NCovar}_{I_3}(\bp_3-\bpt_3)  =  e^{-(\pi d_2/2\lambda_0L)^2 \abs{\bp_3-\bpt_3}^2 },
\end{equation}
which is the normalized covariance for first-order speckle produced by propagation from plane~2 to plane~3.  Indeed, when $\Omega_{01} \gg 1$ and $\Omega_{12} \gg 1$, the law of large numbers implies  
$\langle I_2(\bp_2)\rangle_1 \approx \langle I_2\rangle$ \emph{and} $\langle I_3(\bp_3)\rangle_2 \approx \langle I_3\rangle$, so what would have been speckled speckled speckle in $I_3(\bp_3)$ reduces to the first-order speckle for propagation from plane~2 to plane~3~\cite{footnote5}.

Taking values close to what we might expect in practice---$\lambda_0 = 532\nm$ optical wavelength, $L = 1\m$ to 10\,m scene depth and standoff, $d_0=\ $1\,mm to 1\,cm spot size, $d_1=\ $3\,cm to 2\,m target size, and $d_2=\ $1\,m to 10\,m wall size---we find that $20 \le \W{01} \le 10^9$ and $2\times 10^9 \le \W{12} \le 10^{15}$. The least favorable attenuation factor in Eq.~(\ref{eq:sp3}), $1/(1+\W{01}) \approx 0.05$, is already small enough to make cw third-order speckle reduce to first-order speckle. 

\subsection{Saturation signal-to-noise ratio}
The saturation SNR is related to the normalized covariance $I_3(\bp_3)$ as follows,
\begin{equation}
{\rm SNR}_{\rm sat} = \frac{(\pi D^2/4)^2}{\int_{|\bp_3|\le D/2}\!{\rm d}^2\bp_3\int_{|\tilde{\bp}_3|\le D/2}\!{\rm d}^2\tilde{\bp}_3\,{\rm NCovar}_{I_3}(\bp_3-\tilde{\bp}_3)}.
\end{equation}
Switching to sum and difference coordinates, $\bp_+=\parens{\bp_3+\bpt_3}/2$ and  $\bp_-=\bp_3-\bpt_3$, the $\bpp$ integration yields
\begin{align}
	{\rm SNR}_{\rm sat} = \frac{(\pi D^2/4)^2}{\int_{|\bpm| \le D}{\rm d}^2\bp_-\, \text{NCovar}_{I_3}(\bpm) O(\bpm,D)} 
	\label{SNRsatIntegral}
\end{align}
where
\begin{align}
	O(\bp_-,D) =\frac{\displaystyle D^2}{\displaystyle 2}\bracket{\cos^{-1}\!\parens{\frac{\displaystyle |\bp_-|}{\displaystyle D}}-\frac{\displaystyle |\bp_-|}{\displaystyle D}\sqrt{1-\frac{\displaystyle |\bp_-|^2}{\displaystyle D^2}}}, \text{ for $0 \leq |\bp_-| \leq D$},
\end{align}
is the two-circle overlap function.  An exact evaluation of Eq.~(\ref{SNRsatIntegral}) is tedious but results in 
\begin{align}
{\rm SNR}_{\rm sat}  
	= &\left[ \frac{1}{1+\W{01}}+ \frac{1}{1+\W{12}} + \frac{1}{1+\W{01}+\W{12}} + \frac{d_1^2}{D^2\W{12}}\parens{1+\sqrt{\W{12}}}\parens{1+\frac{1}{1+\W{01}}}\right. \nonumber\\& - \frac{d_1^2 (2+\W{01})}{D^2\sqrt{\W{12}} (1+\W{01})} B\!\parens{\frac{2D^2 \sqrt{\W{12}}}{d_1^2}} - \frac{d_1^2}{D^2\W{12}} B\!\parens{\frac{2D^2 \W{12}}{d_1^2 (1+\W{12})}}\nonumber\\& 
\left. - \frac{d_1^2}{D^2\W{12} (1+\W{01})} B\!\parens{\frac{2D^2 \W{12} (1+\W{01})}{d_1^2 (1+\W{01} + \W{12})}}  \right]^{-1},
\end{align}
where
\begin{align}
	B(x) \equiv e^{-x}[\BI_0(x) + \BI_1(x)],
\end{align}
with $\BI_n$ being the $n$th-order modified Bessel function of the first kind. 

\begin{figure}[hbt]
	\hspace*{0.9in}
	\includegraphics[width=2.75in]{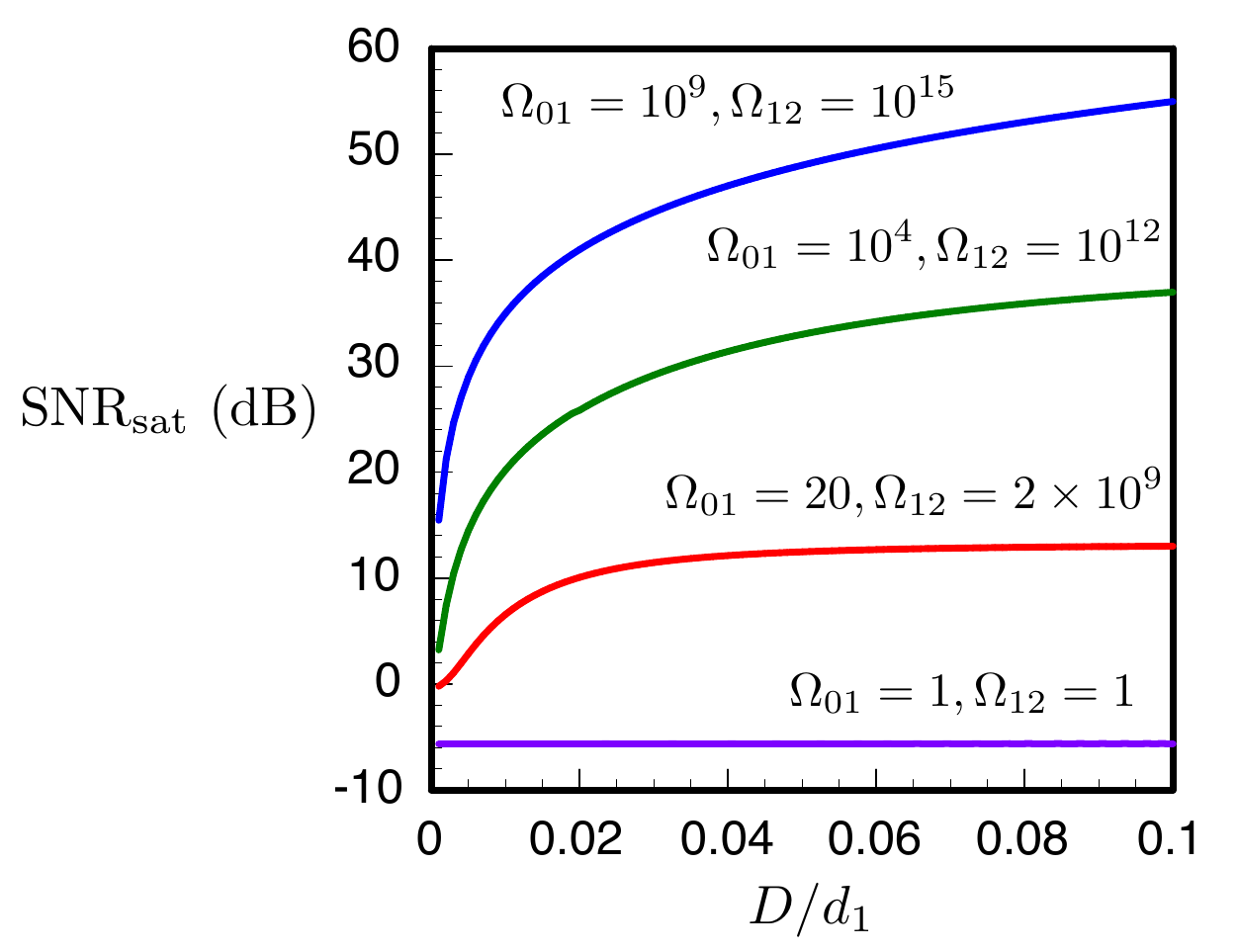}
	\caption{Saturation signal-to-noise ratio, ${\rm SNR}_{\rm sat}$, in dB versus ratio of the detector's diameter to target plane's diameter, $D/d_1$, for four representative $(\Omega_{01},\Omega_{12})$ pairs.
		\label{SNRsat}}
\end{figure}

In Figure~\ref{SNRsat} we have plotted ${\rm SNR}_{\rm sat}$ in dB versus $D/d_1$ for four representative $(\Omega_{01},\Omega_{12})$ pairs:  both being at their maximum values given at the end of Sec.~\ref{irradcovar}, both at their minimum values given there, both at intermediate values between those maxima and minima, and both having unit values.  As a specific example, let $\lambda_0 = 532$\,nm, $L = 1$\,m, $d_0 = 1$\,mm, $d_1 = 10$\,cm, and $d_2 = 1$\,m.  We then have $\Omega_{01} = 2.18 \times 10^4$ and $\Omega_{12}= 2.18 \times 10^{10}$, implying that ${\rm SNR}_{\rm sat} = 31.4$\,dB for $D=1$\,cm.  In short, modest detector diameters will reduce cw speckled speckled speckle's to a miniscule amount within the range of typical Fresnel-number products.  In contrast, the $(\Omega_{01},\Omega_{12}) = (1,1)$ curve in Fig.~\ref{SNRsat} shows the approach to the small-diffusers, ${\rm SNR}_{\rm sat} = 1/7 \approx 0.143$, limit, viz., for unit Fresnel-number products we have ${\rm SNR}_{\rm sat} \approx 0.27$ for $D$ between 1\,mm and 1\,cm.

It should be noted, however, that speckle averaging at the detector is \emph{not} always desirable. This is especially true for Willomitzer~\emph{et al.}'s~\cite{smu} synthetic-wavelength-holography approach to NLoS imaging.  It is a variant of $\scP$-field imaging that: (1) uses sequential, cw illumination at two optical frequencies; (2) heterodyne detects each frequency's $E_3(\bp_3)$ at high spatial resolution, using a detector array, to obtain its speckle pattern; and (3) forms a $\scP$-field image of the hidden-space's target plane by processing the two speckle patterns. Spatial integration over multiple speckles at each detector element degrades the speckle-pattern measurements and is thus undesirable~\cite{Dove2020}. 

\section{Modulated first-order speckle from an extended diffuser \label{ModExtended}}
Our reason for studying modulated speckle is its potential relevance to $\scP$-field NLoS imaging.  With the exception of Teichman's work~\cite{Teichman2019}, prior theoretical treatments of $\scP$-field imaging have ignored the possible ill-effects of high-order speckle on such systems.  The $\scP$ field, as defined in \cite{Dove2019}, is the temporal Fourier transform of the diffuser-averaged STA irradiance,
\begin{align}
	\scP_z(\bp_z,\wm) \equiv \int\!{\rm d}t\, \langle I_z(\bp_z,t)\rangle e^{i\wm t} = \intw{\wp} \langle \scE_z(\bp_z,\wp+\wm/2) \scE_z^*(\bp_z,\wp-\wm/2)\rangle,
\end{align}
where $\scE_z(\bp_z,\omega)\equiv \int {\text{d}t}\, E_z(\bp_z,t) e^{i\omega t}$ and $\langle \cdot\rangle$ denotes ensemble averaging over all relevant diffusers.
To analyze high-order speckle's impact on estimating $\scP_z(\bp_z,\wm)$ from experimental data, we introduce
\begin{align}
	\scPt_z(\bp_z,\wm) \equiv \intw{\wp} \scE_z(\bp_z,\wp+\wm/2) \scE_z^*(\bp_z,\wp-\wm/2),
\end{align}
whose ensemble average equals $\scP_z(\bp_z,\wm)$.  Unfortunately, despite our having found quantitative---indeed favorable---results for the cw third-order speckle from extended diffusers, analysis of the modulated speckle for extended diffusers is far more challenging.  Accordingly, we will limit ourselves to the first-order case, i.e., characterizing the $\scPt$-field fluctuations at plane~1, where the hidden target would be located.  Our goals will be the same as those we set for cw speckle:  determining $\scPt_1(\bp_1,\wm)$'s speckle strength and speckle size.

\subsection{Speckle strength}
The spatial incoherence created by plane 0's diffuser plus the central limit theorem imply that $E_1(\bp_1,t)$ is a zero-mean, complex-valued, Gaussian random process that is completely characterized by its space-time MCF $\langle E_1(\bp_1,t)E_1^*(\tilde{\bp}_1,\tilde{t})\rangle_0$.  It follows that $\mathcal{E}_1(\bp_1,\wm)$ is also a zero-mean, complex-valued, Gaussian random process, and its complete characterization is given by its space-frequency MCF $\langle \mathcal{E}_1(\bp_1,\omega)\mathcal{E}_1^*(\tilde{\bp}_1,\tilde{\omega})\rangle_0$.  Gaussian moment factoring now gives us
\begin{align}
	&\avg{|\scPt_1(\bp_1,\wm)|^2}_0 \nonumber\\&= \intw{\wp} \intw{\wpt} \nonumber\\&  \avg{ \scE_1(\bp_1,\wp+\wm/2) \scE_1^*(\bp_1,\wp-\wm/2) \scE_1^*(\bp_1,\wpt+\wm/2) \scE_1(\bp_1,\wpt-\wm/2)}_0 \nonumber\\
	&=  \intw{\wp} \intw{\wpt} \nonumber\\& \Bigg [ \avg{\scE_1(\bp_1,\wp+\wm/2) \scE_1^*(\bp_1,\wp-\wm/2)}_0 \avg{\scE_1^*(\bp_1,\wpt+\wm/2) \scE_1(\bp_1,\wpt-\wm/2)}_0 \nonumber\\
	&+ \avg{\scE_1(\bp_1,\wp+\wm/2) \scE_1^*(\bp_1,\wpt+\wm/2)}_0 \avg{\scE_1^*(\bp_1,\wp-\wm/2) \scE_1(\bp_1,\wpt-\wm/2)}_0 
	\Bigg].
\end{align}
The first term, after integration, is $|\scP_1(\bp_1,\wm)|^2$. Hence the remaining term, which can be expanded via the Fresnel-diffraction formula, is the $\scPt$ field's variance, viz.,
\begin{align}
	\MoveEqLeft {\rm Var}[\scPt_1(\bp_1,\wm)] =  \avg{|\scPt_1(\bp_1,\wm)|^2}_0 - \abs{\scP_1(\bp_1,\wm)}^2 \nonumber\\
	=& \intw{\wp} \intw{\wpt} \nonumber\\& \avg{\scE_1(\bp_1,\wp+\wm/2) \scE_1^*(\bp_1,\wpt+\wm/2)}_0 \avg{\scE_1^*(\bp_1,\wp-\wm/2) \scE_1(\bp_1,\wpt-\wm/2)}_0 \\
	=& \frac{1}{L^4} \intw{\wp} \intw{\wpt} \intII{\bp_0} \intII{\bpt_0} \nonumber\\& \scE_0(\bp_0,\wp+\wm/2) \scE_0^*(\bp_0,\wpt+\wm/2) \scE_0^*(\bpt_0,\wp-\wm/2) \scE_0(\bpt_0,\wpt-\wm/2) \nonumber\\ &\times e^{i \parens{\wp-\wpt}\parens{\abs{\bp_1-\bp_0}^2-\abs{\bp_1-\bpt_0}^2}/ 2 c L}.
	\label{strength1}
\end{align}
Invoking $E_0(\bp_0,t)$'s being space-time factorable, Eq.~(\ref{strength1}) becomes
\begin{align}
	\MoveEqLeft {\rm Var}[\scPt_1(\bp_1,\wm)]\nonumber\\ =& \frac{1}{L^4} \intw{\wp} \intw{\wpt} \intII{\bp_0} \intII{\bpt_0} \abs{E_0(\bp_0)}^2 \abs{E_0(\bpt_0)}^2 e^{i \parens{\wp-\wpt}\parens{\abs{\bp_1-\bp_0}^2-\abs{\bp_1-\bpt_0}^2}/ 2 c L}\nonumber\\&\times\scS(\wp+\wm/2) \scS^*(\wpt+\wm/2) \scS^*(\wp-\wm/2) \scS(\wpt-\wm/2), 
	\label{eq:sep}
\end{align}
where $\mathcal{S}(\omega) \equiv \int\!{\rm d}t\,S(t)e^{i\omega t}$.  

Equation~(\ref{eq:sep}) does not lend itself to further evaluation, but it does allow us to establish an upper bound on the variance of $\scPt_1(\bp_1,0)$,
\begin{align}
	\MoveEqLeft {\rm Var}[\scPt_1(\bp_1,0)]  \nonumber\\
	=& \frac{1}{L^4} \intw{\wp} \intw{\wpt} \intII{\bp_0} \intII{\bpt_0} \abs{E_0(\bp_0)}^2 \abs{E_0(\bpt_0)}^2 \abs{\scS(\wp)}^2 \abs{\scS(\wpt)}^2 \nonumber\\&\times e^{i \parens{\wp-\wpt}\parens{\abs{\bp_1-\bp_0}^2-\abs{\bp_1-\bpt_0}^2}/ 2 c L}\label{boundbegin}\\
	\leq & \frac{1}{L^4} \intw{\wp} \intw{\wpt} \intII{\bp_0} \intII{\bpt_0} \abs{E_0(\bp_0)}^2 \abs{E_0(\bpt_0)}^2 \abs{\scS(\wp)}^2 \abs{\scS(\wpt)}^2\nonumber\\ 
	=&\abs{\frac{1}{L^2}\intII{\bp_0}\scP_0(\bp_0,0)}^2  = \abs{\scP_1(\bp_1,0)}^2,
	\label{eq:dc-bound}
\end{align}
which shows that the modulated first-order speckle at zero frequency from an extended diffuser is never stronger than ordinary (cw first-order) speckle.

The preceding bound is seemingly at odds with Teichman's analysis~\cite{Teichman2019} for factorable, single-frequency modulation, which finds the modulation-frequency-component speckle to be \emph{stronger} than ordinary speckle. It turns out, however, that this apparent discrepancy is because Teichman's $\scP$-field speckle analysis applies to the modulation-frequency component, whereas our  bound applies to the zero-frequency component.  We will harmonize the two results by assuming
\begin{align}
	S(t) &= e^{-t^2/T^2}\cos(\wmod t) \\
	\scS(\omega) &= \frac{T\sqrt{\pi}}{2} \parens{e^{-T^2(\omega-\wmod)^2/4} + e^{-T^2(\omega+\wmod)^2/4}},
	\label{ModSpectrum}
\end{align}
with $\wmod T \gg 1$, so that $\mathcal{S}(\omega)$ is confined to narrow bandwidths about $\omega = \pm \Omega$. Then, because $|S(t)|^2 = e^{-2t^2/T^2}[1+\cos(2\wmod t)]/2$, we will be concerned with the speckle strengths in $\scPt_1(\bp_1,2\Omega)$ and $\scPt_1(\bp_1,0)$.

The narrowband nature of $\mathcal{S}(\omega)$ allows us to use 
\begin{align}
	\scS(\wp+\wmod) \scS^*(\wpt+\wmod) \scS^*(\wp-\wmod) \scS(\wpt-\wmod) \approx \frac{\pi^2 T^4}{16} e^{-T^2(\wp^2+\wpt^2)/2}
	\label{narrowbandFreq}
\end{align}
in finding ${\rm Var}[\scPt_1(\bp_1,2\Omega)]$ from Eq.~(\ref{eq:sep}), where it will function as a low-pass filter. As a result, we have $\abs{\wp-\wpt}/2cL \ll \wmod/2cL = \pi / \Lambda L$, where $\Lambda=2\pi c/\wmod$ is the modulation wavelength. Then, if $d_0^2<\Lambda L$, so that $E_0(\bp_0) = \sqrt{I_0}\,e^{-4|\bp_0|^2/d_0^2} \approx 0$ for $\abs{\bp_0} > \sqrt{\Lambda L}$, Fresnel diffraction reduces to Fraunhofer diffraction, yielding
\begin{align}
{\rm Var}[\scPt_1(\bp_1,2\wmod)]&= \left(\frac{\pi I_0 T^2}{4L^2}\right)^2\! \intw{\wp} \intw{\wpt} \intII{\bp_0} \intII{\bpt_0} e^{-8(|\bp_0|^2+|\bpt_0|^2)/d_0^2} \nonumber \\[.05in]
&\times e^{-T^2(\wp^2+\wpt^2)/2} e^{-i \parens{\wp-\wpt} \bp_1 \cdot (\bp_0-\bpt_0) / c L} \label{eq:var-gauss-mod} \\[.05in]
& = \frac{\pi P_0^2 T^2}{32 L^4} \sqrt{\frac{1}{1+\parens{d_0 / 2 c L T}^2\abs{\bp_1}^2}},
	\label{eq:var-narrow-illum}
\end{align}
where $P_0 \equiv \pi d_0^2I_0/8$ is the power illuminating plane~0. 
Equation~(\ref{eq:var-narrow-illum}) applies to reasonably practical scenarios, e.g.,  $\sqrt{\Lambda L} = 22\,$cm for $\Lambda = 5\,$cm and $L=1\,$\m, and interestingly is independent of the modulation frequency.  
This result is maximized on axis, where the square-root term vanishes. Say $\abs{\bp_1} < 2L$, as is certainly necessary for paraxial operation. From $d_0^2 < \Lambda L$ it then follows that 
\begin{align}
(d_0/2 c L T)^2\abs{\bp_1}^2 < (2\pi/\wmod T)^2 L/\Lambda \ll 1,
\end{align}
provided  $\wmod T \gg 2\pi\sqrt{L/\Lambda} \approx 28.1$ for $\Lambda=5\cm$ and $L=1\,$m. In this reasonable regime, the square-root term in Eq.~(\ref{eq:var-narrow-illum}) can be neglected entirely so that
\begin{align}
	{\rm Var}[\scPt_1(\bp_1,2\wmod)] = \pi P_0^2 T^2 / 32 L^4.
	\label{varFinal}
\end{align}

Using the assumptions employed thus far, but strengthening $d_0 < \sqrt{\Lambda L}$ to  $d_0 \ll \sqrt{\Lambda L}$, we can show that
\begin{align}
	\abs{\scP_1(\bp_1,2\wmod)}^2
	=& \left(\frac{\pi I_0T^2}{4L^2}\right)^2 \intw{\wp} \intw{\wpt} \intII{\bp_0} \intII{\bpt_0} \nonumber\\& \times e^{-8(|\bp_0|^2 + |\bpt_0|^2)/d_0^2} e^{-T^2(\wp^2+\wpt^2) / 2} e^{-i 2\wmod \bp_1 \cdot (\bp_0-\bpt_0) / c L} \\
	=& \frac{\pi P_0^2 T^2}{32 L^4} e^{-\parens{d_0\wmod / 2 c L}^2\abs{\bp_1}^2}.
	\label{squaredMeanFinal}
\end{align}
Equations~(\ref{varFinal}) and (\ref{squaredMeanFinal}) indicate that the modulation-frequency speckle is as strong as ordinary speckle on axis and is stronger off axis, in complete agreement with Teichman's analysis. However, we can see that in the worst case the increase is only a factor of 
\begin{align}
	e^{(d_0\wmod/ 2 c L)^2|\bp_1|^2} < e^{(2\pi d_0 / \Lambda)^2} \approx 1.08,
\end{align}
for $\abs{\bp_1} < 2L$, $\Lambda = 5\,\text{cm}$, and $d_0 = 2.2\,\text{mm}$. So, Teichman's result for first-order modulated speckle from an extended diffuser is qualitatively correct, but in the paraxial regime that speckle has approximately ordinary strength. It must be emphasized, however, that our analysis for modulated speckle from extended diffusers does \emph{not} extend beyond the first-order case.  

To illustrate our zero-frequency bound for narrowband modulation,  we use
\begin{align}
	\abs{\scS(\wp)}^2 \abs{\scS(\wpt)}^2 \approx \frac{\pi^2 T^4}{16} \parens{e^{-T^2(\wp - \wmod)^2/2} + e^{-T^2(\wp + \wmod)^2/2}} \parens{e^{-T^2(\wpt - \wmod)^2/2} + e^{-T^2(\wpt + \wmod)^2/2} },
\end{align}
in Eq.~(\ref{boundbegin}) and parallel what we just did for the modulation-frequency component.  
We find that
\begin{align}
	{\rm Var}[\scPt_1(\bp_1,0)] = & \frac{\pi P_0^2 T^2}{16 L^4 \sqrt{1+\alpha^2 |\bp_1|^2}}\Bigg( 1 + e^{-\wmod^2 T^2 \alpha^2 |\bp_1|^2 / \parens{1+\alpha^2 |\bp_1|^2}} + 2 e^{-\wmod^2 T^2} \nonumber\\& + 4 e^{-\wmod^2 T^2\parens{2+3\alpha^2 |\bp_1|^2} / 4 \parens{1+\alpha^2 |\bp_1|^2}} \Bigg),
	\label{eq:zero-freq-var-full}
\end{align}
where $\alpha\equiv d_0 / 2 c L T$. Equation~(\ref{eq:zero-freq-var-full}) has its unique maximum on axis, where it takes the value
\begin{align}
	{\rm Var}[\scPt_1(\bzero,0)] = \frac{\pi P_0^2 T^2}{8 L^4} \parens{1 + e^{-\wmod^2 T^2} + 2 e^{-\wmod^2 T^2 / 2}}.
\end{align}
Moreover, we have that
\begin{align}
	\abs{\scP_1(\bp_1,0)}^2 = \frac{\pi P_0^2 T^2}{8 L^4} \parens{1 + e^{-\wmod^2 T^2} + 2 e^{-\wmod^2 T^2 / 2}},
\end{align}
which together with the variance result implies that the zero-frequency speckle from the extended diffuser has ordinary strength on axis and is attenuated off axis, in complete agreement with our bound from (\ref{eq:dc-bound}).

\subsection{Speckle size}
As seen in the cw case for third-order speckle from extended diffusers, $\scPt_1(\bp_1,\wm)$'s variance for the modulated speckle from an extended diffuser is not the sole determinant of whether those speckle fluctuations will severely limit estimating $\scP_1(\bp_1,\wm)$ from experimental data.  The speckle covariance is equally important, if not more so. Deriving modulated first-order speckle's irradiance covariance, however, is more difficult than obtaining its cw counterpart, as we shall soon see.  To start, by using $E_0(\bp_0,t)$'s being space-time factorable, Fresnel-diffraction integrals, and Gaussian moment factoring we can obtain
\begin{align}
	\MoveEqLeft {\rm Covar}[\scPt_1(\bp_1,\wm),\scPt_1(\bpt_1,\wm)]= \avg{\scPt_1(\bp_1,\wm)\scPt^*_1(\bpt_1,\wm)}_0-\scP_1(\bp_1,\wm) \scP^*_1(\bpt_1,\wm) \nonumber\\
	\MoveEqLeft = \frac{I_0^2}{L^4} \intw{\wp}\intw{\wpt}\intII{\bp_0}\intII{\bpt_0} e^{-8(|\bp_0|^2+|\bpt_0|^2)/d_0^2}\nonumber\\ \MoveEqLeft \times \mathcal{S}(\wp+\wm/2)\mathcal{S}^*(\wpt+\wm/2)\mathcal{S}^*(\wp-\wm/2)\mathcal{S}(\wpt-\wm/2)\nonumber\\
	\MoveEqLeft \times e^{i [ \parens{\wo+\wp+\wm/2}\abs{\bp_1-\bp_0}^2 - \parens{\wo+\wpt+\wm/2}\abs{\bpt_1-\bp_0}^2 - \parens{\wo+\wp-\wm/2}\abs{\bp_1-\bpt_0}^2 + \parens{\wo+\wpt-\wm/2}\abs{\bpt_1-\bpt_0}^2 ] / 2 c L}.
\end{align}
The differing spatial-coordinate combinations in the four Fresnel-diffraction exponents makes it impossible to group them in any useful way, regardless of their frequency coefficients. Moreover, expanding these exponents' squares (not shown) does not provide a route to simplification. Furthermore, deleting the exponentials altogether to obtain an upper bound, as we did for ${\rm Var}[\scPt_1(\bp_1,0)]$, is not an option because it removes all spatial dependence, i.e., it suppresses the very behavior we are seeking. 

Despite the preceding difficulties, useful insight into ${\rm Covar}[\scPt_1(\bp_1,2\Omega),\scPt_1(\bpt_1,2\Omega)]$ can be obtained in the narrowband-modulation, $d_0^2 < \Lambda L$ case considered earlier.  
Paralleling the work done there we get 
\begin{align}
	\MoveEqLeft {\rm Covar}[\scPt_1(\bp_1,2\wmod),\scPt_1(\bpt_1,2\wmod)]  = \left(\frac{\pi I_0T^2}{4L^2}\right)^2\! e^{i\wmod\parens{\abs{\bp_1}^2-\abs{\bpt_1}^2}/ c L}\nonumber\\
	&\times  \intw{\wp}\intw{\wpt}\intII{\bp_0}\intII{\bpt_0} e^{-8(|\bp_0|^2+|\bpt_0|^2)/d_0^2}e^{-T^2\parens{\wp^2+\wpt^2}/2} \nonumber\\
	&\times e^{i [-\bp_0\cdot\bp_1\parens{\wo+\wp+\wmod}+\bp_0\cdot\bpt_1\parens{\wo+\wpt+\wmod}+\bpt_0\cdot\bp_1\parens{\wo+\wp-\wmod}-\bpt_0\cdot\bpt_1\parens{\wo+\wpt-\wmod}]/ c L}.
\end{align}
This 6D integral can be evaluated in closed form, resulting in
\begin{align}
{\rm Covar}[\scPt_1(\bp_1,2\Omega),&\scPt_1(\bpt_1,2\Omega)] = \frac{\pi P_0^2T^2}{32 L^4}\frac{e^{i\Omega(|\bp_1|^2-|\bpt_1|^2)/cL}e^{-\alpha^2\Omega^2T^2|\bp_-|^2/4}}{\sqrt{(1+ \alpha^2|\bp_+|^2)(1 + \alpha^2|\bp_-|^2/4) - \alpha^4(\bp_+\cdot\bp_-)^2/4}}
\nonumber \\[.05in]
&\hspace*{.1in}\times \exp\!\left\{-\frac{\wo^2T^2}{4}\frac{\alpha^2|\bp_-|^2+\alpha^4\bracket{|\bp_+|^2|\bp_-|^2-(\bp_+\cdot\bp_-)^2}}{(1+ \alpha^2|\bp_+|^2)(1 + \alpha^2|\bp_-|^2/4) - \alpha^4(\bp_+\cdot\bp_-)^2/4}\right\}, \label{modulatedCovar}
\end{align}
where, as before, $\bp_+ \equiv (\bp_1+\bpt_1)/2$, $\bp_- \equiv \bp_1-\bpt_1$, and $\alpha\equiv d_0/2cLT$.

Despite its formidable length, Eq.~(\ref{modulatedCovar}) readily yields important insights.  First, we have that
\begin{equation}
\lim_{T\rightarrow \infty}\frac{|{\rm Covar}[\scPt_1(\bp_1,2\Omega),\scPt_1(\bpt_1,2\Omega)]|}
{\sqrt{{\rm Var}[\scPt_1(\bp_1,2\Omega)]{\rm Var}[\scPt_1(\bpt_1,2\Omega)]}}  = 
e^{-(\wo^2+\Omega^2)d_0^2|\bp_-|^2/16c^2L^2}.
\label{ModSpeckleSize}
\end{equation}
Second, comparing with Eq.~(\ref{eq:sp1})---and remembering that $\Omega \ll \wo$---shows that modulated first-order speckle has approximately the \emph{same} speckle size as cw first-order speckle in the $T\rightarrow \infty$ limit.  Moreover, because $\max\!\parens{|\bp_+|^2} = d_1^2/4$, $\max\!\parens{|\bp_-|^2} = d_1^2$, $\max\!\bracket{|\bp_+|^2|\bp_-|^2 - (\bp_+\cdot\bp_-)^2} = d_1^4/16$, and $\max\!\bracket{\parens{\bp_+\cdot\bp_-}^2} = d_1^4/64$, this speckle-size near coincidence holds for $T$ satisfying $d_0^2d_1^2/16c^2L^2T^2 \ll 1$.  Taking $d_0 = 3\,$mm, $d_1 =10$\,cm, and $L = 1$\,m we have that $d_0^2d_1^2/16c^2L^2T^2  \le 0.0625$ for $T \ge 1$\,ps, which is much \emph{smaller} than the minimum $T$ value for validating the narrowband ($\Omega T \gg 1$) quasimonochromatic ($\Omega \ll \wo$) modulation assumed in this covariance analysis.  Furthermore, as we have already shown, the first-order speckle strength of the modulated case is close to that of the cw case. Thus, should these behaviors hold for second-order and third-order speckle, the situation would be quite favorable for speckle-suppression in $\scP$-field NLoS imaging, i.e., it would mean that the speckled speckled speckle from extended diffuse reflectors would reduce to the speckle produced by the last reflection \emph{and} that final speckle could easily be averaged out in power collection over realistically-sized detectors.  

\section{Summary and discussion}
Increasing interest in NLoS imaging with coherent illumination is driving the need to understand high-order speckle, principally the speckled speckled speckle that arises in three-bounce NLoS imaging.  This need is especially pressing for $\scP$-field NLoS imaging because initial $\scP$-field experiments have afforded some of the best NLoS coherent-illumination imagery obtained to date~\cite{Liu2019}.  Inasmuch as $\scP$-field imaging relies on \emph{modulated} coherent illumination, the understanding to be sought should encompass modulated speckled speckled speckle.  

In this paper, we took first steps in addressing these issues using space-time factorable initial illumination in a three-diffuser transmissive geometry that is a proxy for three-bounce NLoS imaging.   In the small-diffusers limit we showed that the irradiance variances of cw and modulated $n$th-order speckle coincide and are $(2^n-1)$-times those of ordinary (first-order) speckle.  If not mitigated, this result implies that the \emph{maximum} SNR in three-bounce NLoS imaging, i.e., its ${\rm SNR}_{\rm sat}$, would be 1/7.  The more important case for NLoS imaging, however, involves extended diffuse reflectors.  For our transmissive geometry with extended diffusers, we treated third-order cw speckle and first-order modulated speckle.  It turned out that speckle is unlikely to impede successful operation of coherent-illumination cw imagers because typical parameter values for NLoS scenarios reduce cw third-order speckle in our transmissive-geometry proxy to the ordinary (first-order) speckle produced by the last diffuser. In addition, speckle averaging in optical power collection over typical detector sizes suppresses that residual first-order speckle to the point that its impact on the cw imager's SNR is quite benign, i.e., ${\rm SNR}_{\rm sat} \gg 1$ is achieved.  More importantly, insofar as $\scP$-field NLoS imaging is concerned, our analysis of modulated first-order speckle revealed that its speckle strength and speckle size were very similar to those of the cw case.  So, should the same correspondence apply to modulated third-order speckle, then $\scP$-field NLoS imagers would be largely immune to the adverse effects of speckle.  

It remains to work out the behavior---in speckle strength and speckle size---of modulated third-order speckle for extended diffusers.  Based on what we have accomplished, that important task appears to be quite formidable. A second remaining task of significance is the design and execution of experiments that establish the extent to which speckle effects are actually discernible in coherent-illumination NLoS imaging. 

\section*{Funding}
This work was supported by the DARPA REVEAL program under Contract HR0011-16-C-0030.

\section*{Acknowledgments}
The authors acknowledge fruitful interactions with members of the DARPA REVEAL teams from the University of Wisconsin and Southern Methodist University.  They also thank Dr. Ravi Athale for organizing a valuable workshop on phasor-field imaging, and Dr. Jeremy Teichman for sharing an early version of his analysis of speckle effects in $\scP$-field imaging.

\section*{Disclosures}
The authors declare no conflicts of interest.\\


\begin{thebibliography}{99}
\bibitem{footnote1}The STA irradiance $I_z(\bp_z,t)$ is the power density, averaged over the time interval $[t-T_a, t]$ where $2\pi/\omega_0 \ll T_a \ll 2\pi/\Delta\omega$, that illuminates the transverse coordinates $\bp_z$ in the $z$-plane.
\bibitem{footnote2}For cw illumination $\Delta\lambda$ is infinite, hence in that case we need only assume that the diffuser is rough at the optical wavelength.
\bibitem{footnote3}In the NLoS scenario, both $h_0(\bp_0)$ and $h_2(\bp_2)$ represent the visible wall.  Nevertheless, they can be taken to be statistically independent in our transmissive-geometry proxy if the NLoS imager's visible-wall illumination falls on a different portion of that wall than what lies in its camera's field of view.
\bibitem{footnote4}That the diffuser-averaged STA irradiance is independent of the transverse coordinate is an artifact of Fresnel diffraction and total spatial incoherence produced by our delta-function approximation to $\langle e^{i\omega_0[h_0(\bp_0)-h_0(\tilde{\bp}_0)]/c}\rangle$.  Note that $\langle I_1\rangle < \langle I_0\rangle$ holds because $d_0\ll L$ is required for Fresnel diffraction to be valid.
\end{thebibliography}
\end{document}